\documentclass{iopart}
\usepackage{amsfonts}
\usepackage{amssymb,bbm}
\usepackage[dvips]{graphicx}
\expandafter\let\csname equation*\endcsname\relax
\expandafter\let\csname endequation*\endcsname\relax
\usepackage{amsmath}

\newcommand{\ket}[2][]{{|#2\rangle_{#1}}}
\newcommand{\bra}[2][]{{}_{#1}\langle #2|}
\newcommand{\proj}[2][]{\ket{#2}_{#1}\bra{#2}}
\newcommand{\openone}{\mathbbm 1}
\newcommand{\PolH}{H}
\newcommand{\PolV}{V}

\begin{document}
\title{Experimental generation of complex noisy photonic entanglement}
\author{K~Dobek$^{\textrm{1,2}}$,~M~Karpi{\'n}ski$^{\textrm{3}}$,~R~Demkowicz-Dobrza{\'n}ski$^{\textrm{3}}$, K~Banaszek$^{\textrm{1,3}}$,~P~Horodecki$^{\textrm{4}}$}
\address{$^{\textrm{1}}$ Institute of Physics, Nicolaus Copernicus University, ul.~Grudziadzka 5/7, 87-100 Toru\'{n}, Poland}
\address{$^{\textrm{2}}$ Faculty of Physics, Adam Mickiewicz University, ul.~Umultowska 85, 61-614 Pozna{\'n}, Poland}
\address{$^{\textrm{3}}$ Faculty of Physics, University of Warsaw, Ho\.{z}a 69, 00-681 Warsaw, Poland}
\address{$^{\textrm{4}}$ Faculty of Applied Physics and Mathematics, Technical University of Gda{\'n}sk, 80-952 Gda{\'n}sk, Poland}
\ead{krzysztof.dobek@amu.edu.pl}

%\date{\today}

\begin{abstract}
We present an experimental scheme based on spontaneous parametric down-conversion to produce multiple photon pairs in maximally entangled polarization states using an arrangement of two type-I nonlinear crystals. By introducing correlated polarization noise in the paths of the generated photons we prepare mixed entangled states whose properties illustrate fundamental results obtained recently in quantum information theory, in particular those concerning bound entanglement and privacy.
\end{abstract}

  %DODAC KOD PACS	http://www.aip.org/pacs
\submitto{Laser Phys.}
\maketitle

\section{Introduction}

Recent theoretical studies of noisy entanglement resulted in discoveries of interesting phenomena that occur in higher-dimensional systems. One prominent example is provided by bound-entangled states that cannot be created by local operations and classical communication (LOCC), but at the same time cannot be brought into a pure maximally entangled form under LOCC constraints \cite{HoroHoroPRL98}. Another profound observation is that certain states can be used to generate the cryptographic key at rates that exceed distillable entanglement \cite{HoroHoroPRL05}. These effects reveal the highly complex nature of noisy entanglement in higher dimensions. Experimental studies of such entanglement are important for two main reasons. On the fundamental side, it is always vital to confirm theoretical predictions in actual physical systems. From a more practical perspective, quantum entanglement is an essential resource in implementing a wide range of quantum-enhanced protocols for communication, sensing, etc. However, its generation and distribution are usually affected by imperfections which deteriorate the quality of the available resource. Therefore one needs tools to characterize relevant properties of noisy entanglement and to utilize it for practical purposes in an optimal way.

In this paper we present an experimental scheme that can be used to produce a wide range of noisy entangled multiphoton states. Our approach exploits the geometric symmetry of the popular scheme to generate polarization-entangled photon pairs via spontaneous parametric down-conversion (SPDC) realized in two type-I crystals \cite{KwiaWaksPRA99}. As we describe below, the axial symmetry of this scheme allows one to collect simultaneously several entangled photon pairs with little additional effort. It is worth noting that the symmetry of type-I SPDC has been also exploited in schemes for generating hyperentangled two-photon states \cite{Hyp1,Hyperentanglement} and creating high-dimensional path entanglement of two photons \cite{RossVallPRL09}.

In order to prepare noisy entangled states in the polarization degree of freedom, the generated photons can be subjected to variable correlated birefringence.
We have used this approach to produce two specific examples of four-photon mixed entangled states. One of them is the Smolin state \cite{SmolPRA00}, whose first experimental generation carried out by Amselem and Bourennane \cite{AmseBourNPH09} illustrated the delicate nature of the phenomenon of bound entanglement in the presence of experimental imperfections \cite{LavoKaltNPH2010}. This initial work was subsequently followed by the development of more robust preparation techniques \cite{KampBrusPRA10,LavoKaltPRL10}. Our results indicate that the problems encountered in Ref.~\cite{AmseBourNPH09} are rather generic and do not depend on the specifics of the optical setup. The second example presented here is a private state, for which the secure key contents is strictly higher than the amount of entanglement distillable in the asymptotic regime. These information theoretic properties have been verified experimentally for the first time in \cite{DobeKarpPRL11}. Here we report an experimental illustration of a recent theoretical result which links privacy to incompatibility with local realistic theories \cite{AuguCavaPRL10}. Let us note that non-trivial forms of noisy entanglement have been recently observed also in a system of trapped ions subjected to decoherence induced by spontaneous decay \cite{BarrSchiNPH10}.

This paper is organized as follows. First, we present the experimental setup in Sec.~\ref{Sec:ExpSetup}. The examples of noisy entangled states along with details of their generation are described in Sec.~\ref{Sec:Generation}. Results of their characterization are reviewed in Sec.~\ref{Sec:Characterisation}. Finally, Sec.~\ref{Sec:Conclusions} concludes the paper.

\section{Experimental setup}
\label{Sec:ExpSetup}

Owing to a simple setup and room-temperature operation, SPDC has been the tool of choice to generate multiphoton entangled states. The basic ingredient is an arrangement to produce one maximally entangled photon pair. A very popular configuration is based on a single type-II crystal, with photon pairs collected from the intersection points of two emission cones with orthogonal polarizations \cite{KwiaMattPRL95}. Reflecting the pump pulse and sending it back to the same crystal allows one to generate two pairs. This approach, used to demonstrate quantum teleportation \cite{Teleportation}, has been highly refined in subsequent experiments \cite{PanDaniPRL01}. An alternative is to send the pump pulse through a sequence of type-II crystals \cite{LuZhouNPhysics07}, which enables generation of three or more pairs at once. Furthermore, the alignment can be optimized independently for each crystal.

Another configuration to produce photon pairs in a maximally entangled polarization state utilizes SPDC in two type-I crystals whose optical axes are aligned in perpendicular planes, thus producing orthogonally polarized photons \cite{KwiaWaksPRA99}. Because the down-converted photons are generated as ordinary rays, they emerge on an axially symmetric cone. With a suitably adjusted polarization of the pump beam both the crystals produce photon pairs with the same probability. If the photons generated in the first and the second crystal are indistinguishable apart from their polarization, the source produces a maximally entangled polarization state. An attractive feature of this configuration is that owing to the axial symmetry of the type-I process, several entangled qubit pairs can be collected at once from different locations on the down-conversion cone. This provides a convenient source of multiple photon pairs from a single set of crystals without a need to redirect the pump beam.

\begin{figure}
\begin{center}
\includegraphics[width=10cm]{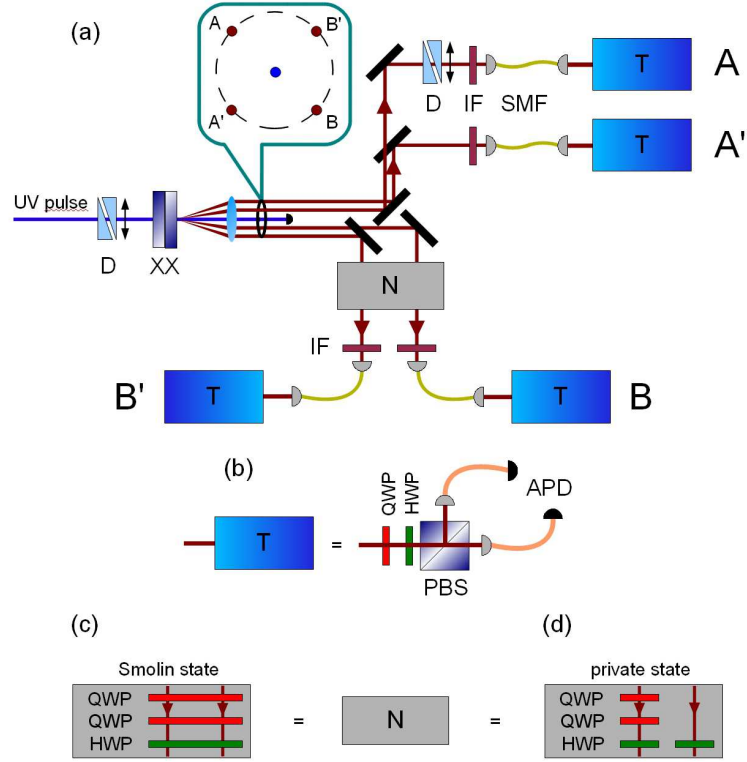}
\end{center}
\caption{(a) A schematic of the experimental setup to produce multiple polarization-entangled photon pairs. Boxes labelled with T are polarization analyzers detailed in (b). The box N represents polarization noise introduced by sets of wave plates shown for the Smolin state (c) and the private state (d).
D, Soleil-Babinet compensator; XX, down-conversion crystsls; IF, interference filter; SMF, single-mode fiber; QWP, quarter-wave plate; HWP, half-wave plate; PBS, polarizing beam splitter; APD, avalanche photodiode.}
\label{Fig:Setup}
\end{figure}

We implemented the above idea in a setup shown schematically in Figure~\ref{Fig:Setup}. The pump beam was a $78~\mathrm{MHz}$ train of
$180~\mathrm{fs}$ pump pulses with the spectrum centered at $390~\mathrm{nm}$, obtained by frequency doubling in a $1~\mathrm{mm}$ long lithium triborate crystal the output from a Ti:sapphire oscillator (Coherent Chameleon Ultra), which resulted in $200~\mathrm{mW}$ average power. The pump beam was focused to a $70~\mu\mathrm{m}$ diameter spot in the pair of beta-barium borate crystals. The SPDC emission was collimated with a 20~cm focal length lens, sending the generated photons along parallel paths. The idea of this arrangement is similar to that used recently to generate multipath entanglement of photon pairs \cite{RossVallPRL09}.

In the constructed setup, it was possible to introduce various types of polarization noise by inserting in the paths of one or two photons birefringent elements (half- and quarter-wave plates) with variable orientations. Two Soleil-Babinet compensators were placed in the pump beam and the path of one of the down-converted photons to control relative phases between the horizontal and vertical components for both the produced pairs. In experiments where photon polarizations were measured individually the photons were filtered through 10~nm interference filters and coupled into single-mode fibers, which delivered them to polarization analyzers shown in Figure~\ref{Fig:Setup}(b). Each analyzer consisted of a quarter- and half wave plate followed by a Wollaston polarizing beam-splitter, whose output ports were coupled into multimode fibers. These fibers guided the photons to avalanche photodiodes operated in the Geiger mode (Perkin-Elmer SPCM-AQR). The electronic signals from the detectors were finally processed using a coincidence circuit with a 6~ns window programmed in an field-programmable gate array (FPGA) board. All the waveplates were mounted on motorized rotation stages and the whole experiment was controlled using a dedicated LabView (National Instruments) application. Typical count rates in the setup were of the order of $10^5$~Hz for single counts, $10^4$~Hz for two-fold coincidences for detectors monitoring the same photon pair, and $2$~Hz for four-fold coincidences.

\section{Generation of noisy states}
\label{Sec:Generation}

We employed the setup described in the preceding section to generate and characterize noisy four-qubit states that illustrate interesting phenomena occurring in the theory of high-dimensional entanglement. The first example was the Smolin state \cite{SmolPRA00} of four qubits $ABA'B'$. It is defined as an equally weighted statistical mixture of four components, each corresponding to both the pairs prepared in the same Bell state:
\begin{multline}
\hat{\varrho}_{\text{Smolin}}
= \frac{1}{4} \bigl( \proj[AB]{\phi_+} \otimes \proj[A'B']{\phi_+} +  \proj[AB]{\phi_-} \otimes \proj[A'B']{\phi_-} \\
+ \proj[AB]{\psi_+} \otimes \proj[A'B']{\psi_+} +  \proj[AB]{\psi_-} \otimes \proj[A'B']{\psi_-}
\bigr).
\label{Eq:SmolinDef}
\end{multline}
We denoted here the Bell states as:
\begin{eqnarray}
\ket{\phi_\pm} & = \frac{1}{\sqrt{2}} \bigl( \ket{\PolH\PolH} \pm \ket{\PolV\PolV} \bigr) \nonumber \\
\ket{\psi_\pm} & = \frac{1}{\sqrt{2}} \bigl( \ket{\PolH\PolV} \pm \ket{\PolV\PolH} \bigr),
\end{eqnarray}
where $H$ and $V$ stand for the horizontal and the vertical polarization respectively.
The Smolin state $\hat{\varrho}_{\text{Smolin}}$ is invariant with respect to any permutation of individual qubits, which can be seen most easily from its representation in terms of Pauli matrices $\hat{\sigma}^\mu$:
\begin{equation}
\hat{\varrho}_{\text{Smolin}} = \frac{1}{16} \left( \vphantom{\sum_{\mu=x,y,z}}
\hat{\openone}_{A} \otimes \hat{\openone}_{B} \otimes \hat{\openone}_{A'} \otimes \hat{\openone}_{B'} + \sum_{\mu=x,y,z} \hat{\sigma}^{\mu}_{A} \otimes \hat{\sigma}^{\mu}_{B} \otimes
\hat{\sigma}^{\mu}_{A'} \otimes  \hat{\sigma}^{\mu}_{B'}
\right),
\end{equation}
where we used the standard notation
\begin{eqnarray}
\hat\sigma^x & =\ket H \bra V+\ket V \bra H \nonumber \\
\hat\sigma^y & =\rmi \bigl(\ket V \bra H-\ket H \bra V \bigr) \nonumber \\
\hat\sigma^z & = \ket H \bra H-\ket V \bra V.
\end{eqnarray}

If we consider a partition of the four qubits into two subsystems, the first one comprising just one qubit and the second one remaining three, e.g.\ $A\mathop{:}BA'B'$, the Smolin state exhibits bipartite entanglement. Indeed, the Bell state measurement applied by the second party to qubits $A'B'$ allows her to bring the qubits $AB$ to a maximally entangled state without communication with the other party. On the other hand, the state is separable with respect to any partition into two pairs of qubits, which follows immediately from the construction of the state as a statistical mixture specified in Eq.~(\ref{Eq:SmolinDef}) and permutational invariance. This implies that if each qubit is in hands of a separate party and all the parties are restricted to LOCC manipulations only, then it is impossible to distill any entanglement. Thus the Smolin state is an example of bound entanglement \cite{HoroHoroPRL98}.

To generate the Smolin state in our experimental setup, we started from two maximally entangled pairs of photons $AB$ and $A'B'$. With suitable settings of the Soleil-Babinet compensators, both the pairs were prepared in the same state $\ket{\phi_+}$. This can be converted into the Smolin state by applying randomly to one qubit from each pair, chosen to be $B$ and $B'$, equiprobable transformations $\hat{\openone}_B \otimes \hat{\openone}_{B'}$, $\hat{\sigma}^{x}_{B} \otimes \hat{\sigma}^{x}_{B'}$, $\hat{\sigma}^{y}_{B} \otimes \hat{\sigma}^{y}_{B'}$, or $\hat{\sigma}^{z}_{B} \otimes \hat{\sigma}^{z}_{B'}$. Differently from Ref.~\cite{AmseBourNPH09}, we realized these transformations by sending the photons $B$ and $B'$ through the same set of three waveplates, as depicted in Figure~\ref{Fig:Setup}(c).
Two quarter-wave plates are equivalent either to the identity transformation or a half-wave plate depending on whether their fast axes are mutually parallel or perpendicular. Further, a suitably oriented half-wave plate realizes $\hat\sigma^x$ or $\hat\sigma^z$. This allowed us to implement all four transformations required to generate the Smolin state, since $\hat{\sigma}^y = \rmi \hat\sigma^x \hat\sigma^z$.

The second example of noisy entanglement which we produced experimentally was a four qubit state
\begin{multline}
\hat{\varrho}_{\text{private}}
= \frac{1}{4} \bigl( \proj[AB]{\phi_-} \otimes \proj[A'B']{\psi_-} +  \proj[AB]{\psi_+} \otimes \proj[A'B']{\phi_+} \\
+ \proj[AB]{\psi_+} \otimes \proj[A'B']{\psi_+} +  \proj[AB]{\psi_+} \otimes \proj[A'B']{\phi_-}
\bigr).
\label{Eq:PrivateDef}
\end{multline}
This state has nontrivial properties in the context of quantum key distribution and therefore we refer to it as the private state. Let us assume that Alice and Bob are respectively in possession of qubits $AA'$ and $BB'$. Suppose that they measure qubits $A$ and $B$ in the eigenbasis of the $\hat{\sigma}^y$ operator composed of two vectors $\ket{\bar{0}}$ and $\ket{\bar{1}}$ defined as
\begin{equation}
\ket{\bar{\upsilon}} = \frac{1}{\sqrt{2}} \bigl( \ket{H} + \rmi (-1)^\upsilon\ket{V}\bigr), \qquad \upsilon=0,1.
\end{equation}
The reduced density matrix of the qubits $A$ and $B$ expressed in this basis takes the form
\begin{equation}
\Tr_{A'B'}(\hat{\varrho}_{\text{private}}) = \frac{1}{2} \bigl( \proj[AB]{\bar{0}\bar{0}}
+ \proj[AB]{\bar{1}\bar{1}}\bigr)
- \frac{1}{4} \bigl( \ket[AB]{\bar{0}\bar{0}} \bra{\bar{1}\bar{1}}
+ \ket[AB]{\bar{1}\bar{1}} \bra{\bar{0}\bar{0}} \bigr).
\label{Eq:TrABrhoprivate}
\end{equation}
It is clearly seen that measurements performed by Alice and Bob
on the qubits $A$ and $B$ in the $\hat{\sigma}^y$ eigenbasis yield equiprobable and perfectly correlated results $0$ or $1$. One may ask whether these results form a secure cryptographic key. If Alice and Bob had access only to qubits $A$ and $B$ this would not be the case, as the magnitude of the off-diagonal elements in Eq.~(\ref{Eq:TrABrhoprivate}) is less than $\frac{1}{2}$, which means that $\Tr_{A'B'}(\hat{\varrho}_{\text{private}})$ is not maximally entangled. However, the presence of qubits $A'$ and $B'$ turns out to guarantee perfect security. As discussed in \cite{HoroHoroPRL05}, these two additional qubits serve as the {\em shield subsystems} preventing an eavesdropper from accessing any information about measurement results. In this context, $A$ and $B$ are often referred to as {\em key subsystems}. Interestingly, the four-qubit state $\hat{\varrho}_{\text{private}}$ has distillable entanglement strictly less than one, which means that asymptotic conversion into a smaller number of maximally entangled singlet pairs is not the optimal way to generate a cryptographic key from $\hat{\varrho}_{\text{private}}$.

The state $\hat{\varrho}_{\text{private}}$ can be generated using our source of multiple photon pairs using an arrangement of wave plates shown in Figure~\ref{Fig:Setup}(d). The photon $B$ is sent trough a half-wave plate which realizes either $\hat{\sigma}^x$ or $\hat{\sigma}^z$, while the photon $B'$ travels through a set of three wave plates that can be set to implement identity or any Pauli operator. Applying these transformation in a correlated manner as $\hat{\sigma}^{z}_{B} \otimes \hat{\sigma}^{y}_{B'}$, $\hat{\sigma}^{x}_{B} \otimes \hat{\openone}_{B'}$, $\hat{\sigma}^{x}_{B} \otimes \hat{\sigma}^{x}_{B'}$, and $\hat{\sigma}^{x}_{B} \otimes \hat{\sigma}^{z}_{B'}$ yields the private state. Its generation and experimental analysis of information-theoretic properties has been first reported in \cite{DobeKarpPRL11}.

\section{Experimental characterization}
\label{Sec:Characterisation}

We performed a full tomographic reconstruction of the four-qubit state by sending individual photons to polarization analyzers and measuring all $81$ combinations of projections in the eigenbases of the operators $\hat\sigma^x$, $\hat\sigma^y$, and $\hat\sigma^z$. For each combination, four-fold coincidences were recorded over an interval of approximately 1~hour, resulting in the total time of an experimental run equal to about 81 hours.
The orientation of wave plates introducing polarization noise was changed at $30$~s intervals. The counting circuit was
put on hold during the operation of motors rotating the wave plates which took approximately $10$~s each time.

In order to test the operation of the setup, we adopted a faster alternative procedure to collect data. In this procedure, the FPGA board was programmed to record events composed of pairs of two-fold coincidences triggered between paths $AB$ and $A'B'$, but not necessarily within the time window of the same pulse. Specifically, after registering a two-fold coincidence for one combination of paths (either $AB$ or $A'B'$) the counting circuit waited for a two-fold coincidence between detectors monitoring the other combination, and recording the result as a four-fold event. Additional two-fold coincidences involving the first combination of paths that occurred during the waiting window were ignored. This procedure allowed us to collect four-fold events at a rate that was approximately half the rate of producing single photon pairs, reducing the overall measurement time to approximately $3$ hours. In this case, the limiting factor was the speed of the motorized rotation stages. Data collected using the fast procedure were used to verify the properties of the generated state before a full experimental run.

\begin{figure}
\begin{center}
\includegraphics[width=10cm,bb = 90 10 470 375]{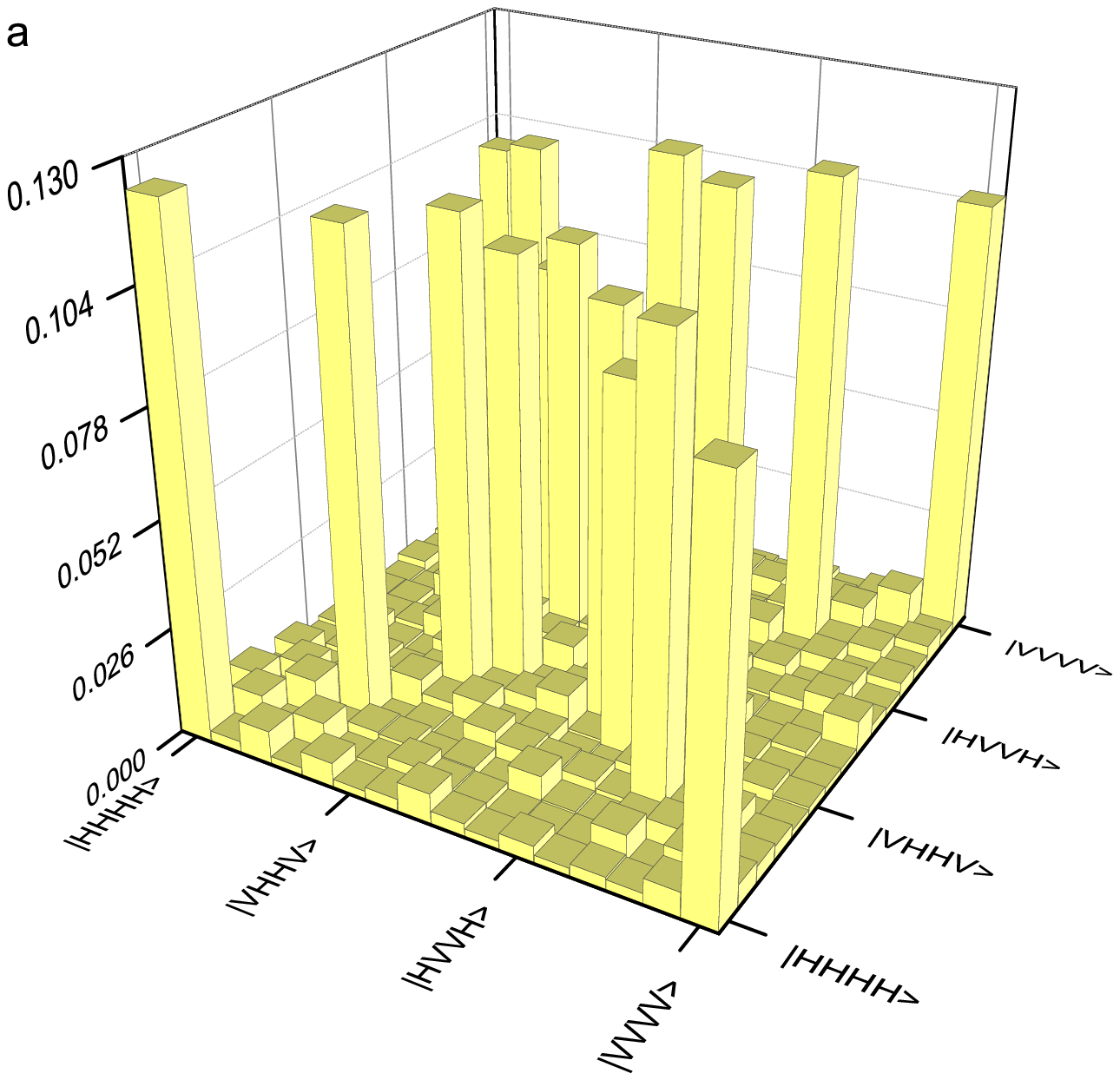}

\vspace{0.125in}

\includegraphics[width=10cm,bb = 55 10 450 385]{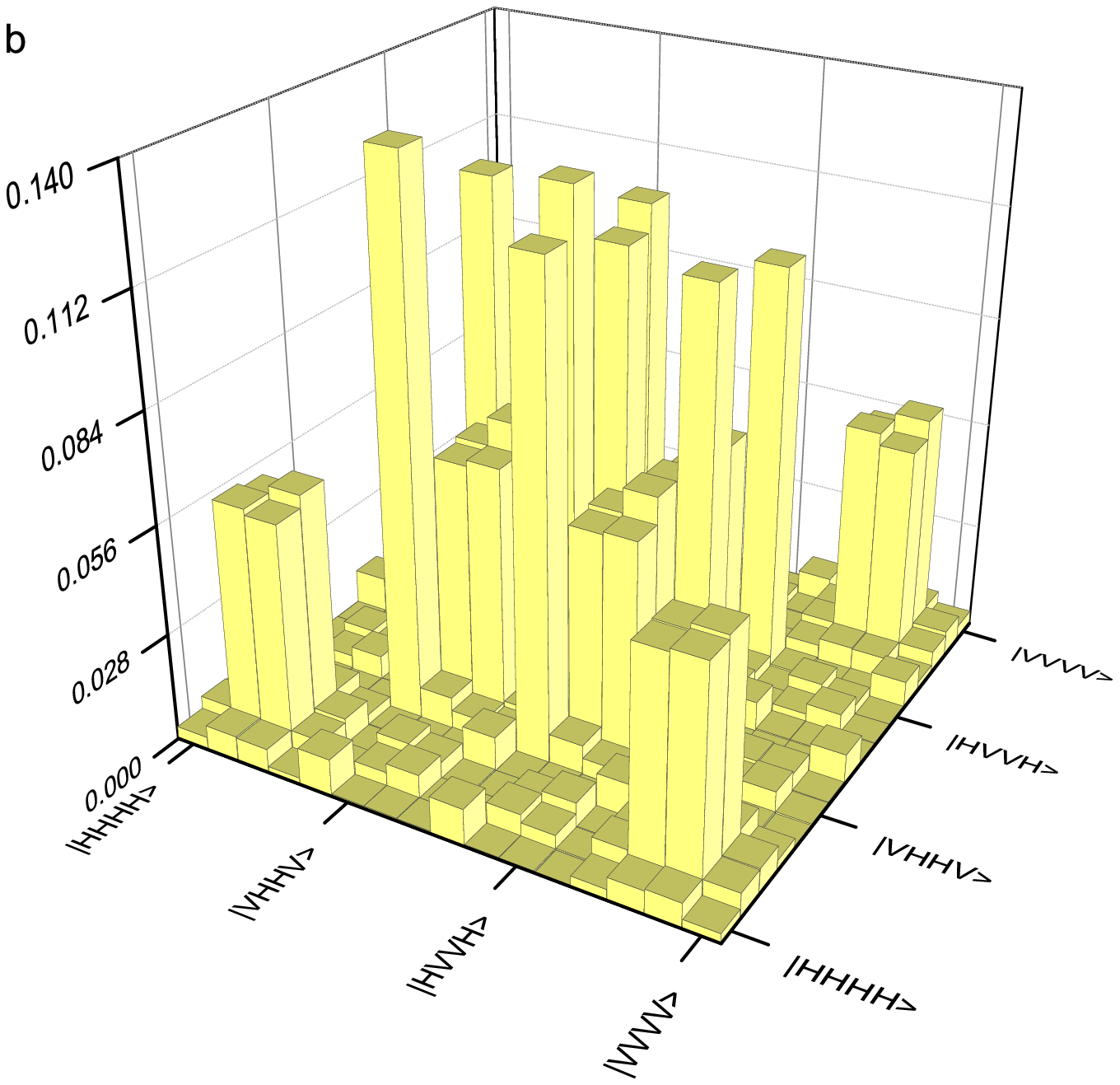}
\end{center}
\caption{Absolute values of the elements of reconstructed density matrices for (a) the Smolin state and (b) the private state. }
\label{Fig:Rho}
\end{figure}

In Figure~\ref{Fig:Rho} we present density matrices for the Smolin state and the private state reconstructed from the experimental data using the maximum likelihood method \cite{HradPRA97,BanaDAriPRA99,JameKwiaPRA01}. These experimental results can be compared against the idealized states by calculating the corresonding fidelities, defined in general for two states $\hat{\varrho}$ and $\hat{\varrho}'$ as
\begin{equation}
F(\hat{\varrho}, \hat{\varrho}') = \Tr \left( \sqrt{\sqrt{\varrho}\varrho'\sqrt{\varrho}}\right).
\end{equation}
We obtained values $F=0.923 \pm 0.002$ for the Smolin state and $F=0.971 \pm 0.001$ for the private state. The experimental Smolin state fidelity is comparable to that obtained in two other experiments aiming at generating $\hat{\varrho}_{\text{Smolin}}$ in a four-photon system \cite{AmseBourNPH09,LavoKaltPRL10} which suggests that it may be at the limit of what can be achieved in a typical realization with standard optical elements. The higher fidelity for the private state may be attributed to the fact that in this case the polarization noise was introduced with fewer birefringent elements, thus reducing the effects of their imperfections.

For the Smolin state, we used polarization measurements on individual photons to determine the expectation value of an entanglement witness given by \cite{AmseBourNPH09}
\begin{equation}
\hat{W} = \hat{\openone}^{\otimes 4} + ( \hat{\sigma}^x )^{\otimes 4}
+ ( \hat{\sigma}^y )^{\otimes 4}
+ ( \hat{\sigma}^z )^{\otimes 4}
\end{equation}
which was found to be equal to $\langle \hat{W} \rangle = -1.43 \pm 0.02$, verifying the nonclassical character of the generated state. We also calculated eigenvalues of the partially transposed density matrix with respect to three possible partitions into two pairs of qubits. The results are presented in Table~\ref{Tab:Eigenvalues}. It is seen that for each partition some of the eigenvalues are negative. A similar feature was also present in the first experimental generation of the Smolin state reported in \cite{AmseBourNPH09}. The reason behind the occurrence of negative eigenvalues is that the theoretical Smolin state $\hat{\varrho}_{\text{Smolin}}$ is located exactly on the boundary of positive partial transposition states. Non-ideal implementation of the polarization noise, imperfect alignment of the polarization analyzers, and the statistical uncertainty of the measured density matrix may therefore easily produce residual entanglement in the reconstructed data for any partition. This problem can be solved by generating a mixture of $\hat{\varrho}_{\text{Smolin}}$ and a completely mixed four-qubit state \cite{KampBrusPRA10,LavoKaltPRL10} which for a suitable choice of relative weights demonstrates the phenomenon of bound entanglement in an experimentally robust way.

\begin{table}
\caption{Eigenvalues of the partially transposed density matrix characterizing the experimentally generated Smolin state, obtained for three possible
partitions into two pairs of qubits. }
\label{Tab:Eigenvalues}
\begin{center}
  \begin{tabular}{  c  c  c  c }
    \hline\hline
    	Theory & $AB\mathop{:}A'B'$ & $AB'\mathop{:}A'B$ & $AA'\mathop{:}BB'$ \\ \hline
	$0.250$ & $0.229$ & $0.228$ & $0.229$ \\
	$0.250$ & $0.216$ & $0.216$ & $0.217$ \\
	$0.250$ & $0.214$ & $0.215$ & $0.213$ \\
	$0.250$ & $0.202$ & $0.202$ & $0.204$ \\
	$0.000$ & $0.036$ & $0.034$ & $0.034$ \\
	$0.000$ & $0.026$ & $0.029$ & $0.029$ \\
	$0.000$ & $0.024$ & $0.025$ & $0.023$ \\
	$0.000$ & $0.022$ & $0.023$ & $0.022$ \\
	$0.000$ & $0.016$ & $0.016$ & $0.015$ \\
	$0.000$ & $0.011$ & $0.011$ & $0.012$ \\
	$0.000$ & $0.008$ & $0.009$ & $0.009$ \\
	$0.000$ & $-0.005$ & $-0.007$ & $-0.006$ \\
	$0.000$ & $-0.003$ & $0.005$ & $0.004$ \\
	$0.000$ & $0.003$ & $-0.004$ & $-0.003$ \\
	$0.000$ & $0.001$ & $-0.002$ & $-0.002$ \\
	$0.000$ & $0.000$ & $0.001$ & $0.001$ \\
    \hline\hline
  \end{tabular}
\end{center}
\end{table}

Experimental characterization of the privacy properties of the state $\hat{\varrho}_{\text{private}}$ has been described in Ref.~\cite{DobeKarpPRL11}. The reconstruction of information-theoretic quantities from experimental data was found to be very sensitive to statistical uncertainties due to highly non-linear dependence on the elements of the density matrix. The statistical distributions for the quantities of interest were obtained by evaluating them on individual density matrices that formed an ensemble consistent with experimental data. The results demonstrated a statistically significant separation between the distillable entanglement and the key contents for the generated state.

Private states exhibit other specifically nonclassical properties. In particular, Augusiak {\em et al.} \cite{AuguCavaPRL10} have recently presented a general theoretical proof that perfect privacy implies incompatibility with local realistic theories. This motivated us to test whether the experimentally generated state, despite its non-ideal privacy, can be used to demonstrate a violation of Bell's inequalities. The proof presented in Ref.~\cite{AuguCavaPRL10} is based on an observation that any private state can be brought by local operations (without classical communication to satisfy the locality requirement) to a form in which the key subsystems exhibit correlations violating local realism. In our case, it is easy to see that for the ideal state $\hat{\varrho}_{\text{private}}$ defined in Eq.~(\ref{Eq:PrivateDef}) no local operations are necessary. This is because the reduced density matrix of the qubits $AB$, given explicitly by
\begin{equation}
\label{Eq:rhoab}
\Tr_{A'B'} (\hat{\varrho}_{\text{private}}) = \frac{1}{4} \proj[AB]{\phi_-} + \frac{3}{4} \proj[AB]{\psi_+},
\end{equation}
is a statistical mixture of two Bell states with unequal weights. With a suitable choice of projective measurements, any such a mixture violates the Clauser-Horne-Shimony-Holt (CHSH) inequality, which follows from the set of necessary and sufficient conditions derived in \cite{HoroHoroPLA95}.

In order to test the CHSH inequality we performed polarization measurements on the key photons in coincidence with detectors monitoring the shield photons. The CHSH inequality can be written as
\begin{equation}
-2 \le {\cal B} \le 2,
\end{equation}
where
\begin{equation}
{\cal B} = C({\bf a}; {\bf b}) + C({\bf a}'; {\bf b}) + C({\bf a}; {\bf b}') - C({\bf a}'; {\bf b}')
\end{equation}
is a combination of four correlation functions
\begin{equation}
C({\bf a}; {\bf b}) = \bigl\langle ({\bf a}\cdot \hat{\boldsymbol\sigma} )\otimes
({\bf b}\cdot \hat{\boldsymbol\sigma} ) \bigr\rangle
\end{equation}
for Bloch vectors ${\bf a}$, ${\bf a}'$, ${\bf b}$, and ${\bf b}'$ that define the measurement bases.
For the specific state given in Eq.~(\ref{Eq:rhoab}), the maximum violation of the CHSH inequality is obtained for the choice of vectors:
\begin{equation}
{\bf a} = \begin{pmatrix} 0 \\ 1 \\ 0 \end{pmatrix}, \quad
{\bf a}' = \begin{pmatrix} 0 \\ 0 \\ 1 \end{pmatrix}, \quad
{\bf b} = \begin{pmatrix} 0 \\ \frac{2}{\sqrt{5}} \\ - \frac{1}{\sqrt{5}}  \end{pmatrix}, \quad
{\bf b}' = \begin{pmatrix} 0 \\ \frac{2}{\sqrt{5}} \\ \frac{1}{\sqrt{5}} \end{pmatrix},
\end{equation}
resulting in the value of the CHSH combination equal to ${\cal B} = \sqrt{5} \approx 2.236$. Polarization measurements performed on
the key photons in these bases yielded the result ${\cal B} = 2.12 \pm 0.01$, which is clearly above the limit permitted by local realistic theories. It is interesting to note that the reduced density matrix of the key qubits $AB$ has been found in Ref.~\cite{DobeKarpPRL11} to contain no distillable key due to experimental imperfections. Thus the violation of the CHSH inequality turns out to be a more robust way to detect quantum correlations contained in this two-qubit state. This is  understandable, since the CHSH combination can be written as a single quantum mechanical observable, while the calculation of the key contents requires highly nonlinear processing of the reconstructed density matrix. This problem may be alleviated by the development of more efficient methods to characterize the amount of distillable key based on a single or a few observables, such as those recently presented in \cite{BanaHoroXXX11}.

\section{Conclusions and outlook}
\label{Sec:Conclusions}

In conclusion, we presented an arrangement to collect photon pairs in maximally entangled polarization states from a single set of two type-I down-conversion crystals. Application of correlated noise introduced using rotating wave plates enabled us to produce noisy entangled four-photon states that illustrated fundamental results of the entanglement theory.

\begin{figure}
\begin{center}
\includegraphics[width=10cm]{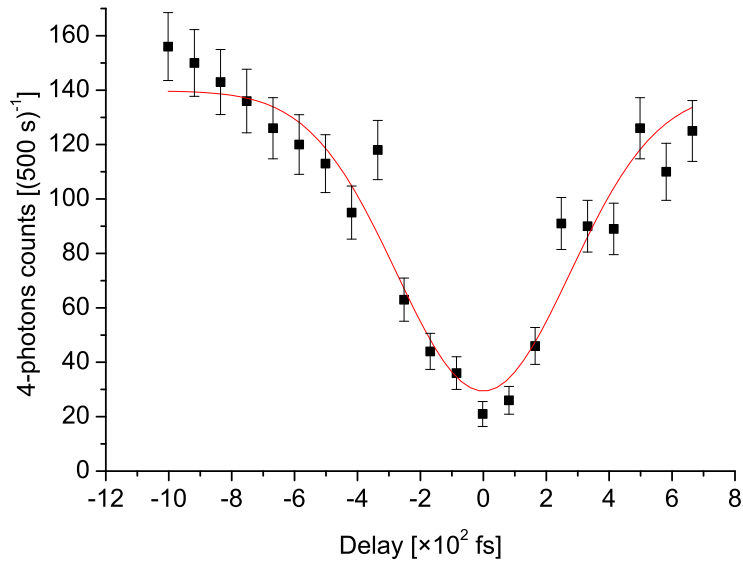}
\end{center}
\caption{The Hong-Ou-Mandel interference dip between heralded photons from two independent pairs generated in a single beta-barium borate crystal. The interfered photons were  transmitted through $2~\mathrm{nm}$  interference filters. The delay was introduced by translating one of the collimators coupling photons into a single-mode fiber. The experimental data (squares with error bars) are fitted with a Gaussian profile (solid line).}
\label{fig:hom}
\end{figure}

Characterization of information-theoretic properties of the generated states was performed by measuring polarizations of individual photons in suitably selected bases. If measurements are restricted to this class, the produced photon pairs can be correlated in other degrees of freedom, e.g.\ frequency, provided that the modal structure of the pair is independent of the polarization state \cite{URenBanaQIC03,AlfredLasPhys}. For the configuration based on two type-I crystals, this condition is satisfied over relatively large bandwidths of the generated photons, which follows from the symmetry of the type-I down-conversion process \cite{DragPRA04}. This allows one to avoid heavy spectral filtering and consequently offers increased four-fold coincidence rates, which are notoriously low in most multiphoton experiments. However, if independently generated photons are to be interfered, it is necessary to ensure their spectral indistinguishability. This requirement can be verified with the help of the Hong-Ou-Mandel two-photon interference effect \cite{HongOuPRL85}. We carried out a preliminary test by producing two pairs in a single crystal and interfering photons from different pairs in an event-ready manner \cite{ZukoZeilPRL93} using a single-mode fiber optic directional coupler with a $50\mathop{:}50$ splitting ratio. When the interfered photons were transmitted through 2~nm bandwidth interference filters, $79\%$ depth of the Hong-Ou-Mandel dip was observed, as shown in Figure~\ref{fig:hom}. This allows for some optimism about using the described source in more sophisticated experiments utilizing multiphoton interference effects.

\ack

We wish to acknowledge insightful discussions with Czes{\l}aw Radzewicz and Wojciech Wasilewski. This research was supported
by FP7 FET projects CORNER (contract no.\ 213681) and Q-ESSENCE (contract no.\ 248095), and the Foundation for Polish Science TEAM project cofinanced by the EU European Regional Development Fund. PH is partially supported by Polish Ministry of Science and Higher
Education grant no. N N202 261938.

\end{document}